\documentclass[epj,referee]{svjour}
\usepackage{bm}
\usepackage{graphicx}

\begin{document}
\title{The Complexity of the Spherical $p$-spin spin glass model, revisited}

\author{A. Crisanti \and L. Leuzzi \and T. Rizzo}
%\email{andrea.crisanti@phys.uniroma1.it}
%\email{luca.leuzzi@roma1.infn.it}
%\email{tommaso.rizzo@phys.uniroma1.it}

\institute{Dipartimento di Fisica, Universit\`a di Roma ``La Sapienza''\\
           Istituto Nazionale Fisica della Materia, Unit\`a di Roma, 
             and SMC,\\
             P.le Aldo Moro 2, I-00185 Roma, Italy
          }

\date{Received: date / Revised version: date}

\abstract{
Some questions concerning the calculation of the number of ``physical''
(metastable) states or complexity of the spherical $p$-spin spin glass model 
are reviewed and examined further. 
Particular attention is focused on the general calculation procedure 
which is discussed step-by-step.
} 

\PACS{ {75.10.Nr}{Spin-glass and other random models} \and
       {02.30.Mv}{bho}
     }
% end of PACS codes } 
%end of abstract
%

\maketitle

\section{introduction}
\label{sec:Intro}
\noindent
The analysis of the equilibrium and non-equilibrium properties 
in terms of the energy landscape originally pushed forward for the structural 
glass transition \cite{StiWeb82}, has risen in the recent years a new 
interest on 
the topological properties of the energy or free-energy landscapes
of disordered and complex systems.
In this approach an important theoretical tool is the logarithm of the number 
metastable states, called {\it complexity} or {\it configurational entropy},
identified as basins or valleys on the landscape.

Solvable models, such as mean-field models,  
have always played an 
important role in the theoretical study of physical problems. 
In this context classical calculations \cite{BraMoo80,DedYou83}
for the complexity of the Sherrington-Kirkpatrick \cite{SheKir75} (SK) 
and other disordered spin models have been reconsidered, 
extended and in some cases
criticised \cite{CavGiaParMez03,BraMoo03,CriLeuParRiz03}.

Motivated by these criticisms in this paper some questions concerning the 
calculation of the complexity of disordered spin systems are reviewed and 
examined further in a solvable model, 
the spherical $p$-spin spin glass ($p$SP-SG) model introduced by 
Crisanti and Sommers \cite{CriSom92}.
The aim of the paper is not the calculation of the complexity for
the spherical $p$SP-SG model, which has been computed in
Refs. \cite{CriSom95,CavGarGia98,CavGiaPar98}, but the procedure of
calculation itself to clarify know results which could be useful for
the understanding of the complexity of other systems.
The spherical $p$SP-SG model is only used to
enlighten subtle points of the procedure. We have tried to make 
the paper almost self-contained so that it can be also used 
by readers interested into the problem but not too 
familiar with all reported results. 

The general approach to the calculation of complexity in 
mean-field spin glass model is discussed in Section \ref{sec:Comp}.
The procedure is illustrated in 
Section \ref{sec:ComppSP} using the spherical $p$SP-SG model introduced in
Section \ref{sec:pTAP}.
The of correctness the procedure is discussed in Section \ref{sec:HessEig}. 
Finally Section \ref{sec:Disc} contains some conclusions and discussion.

\section{How to compute the Complexity}
\label{sec:Comp}
\noindent
The metastable states in mean-fields spin models are more easily studied 
using the Thouless-Anderson-Palmer (TAP) method \cite{ThoAndPal77}, which 
introduces a mean-field free energy functional $F_{\rm TAP}(\bm{m})$
of the local magnetizations $\bm{m} = (m_1,$ $ m_2, \ldots, m_N)$,
where $N$ is the number of spins. At any temperature $T$
the metastable states of the system are identified with the local minima of 
$F_{\rm TAP}(\bm{m})$, i.e., with the solutions of
\begin{equation}
\label{eq:TAPeq}
 \partial_{m_i} F_{\rm TAP}(\bm{m}) = 0, \qquad i = 1,\ldots,N
\end{equation}
with the additional requirement that {\it all} eigenvalues of the
matrix $\partial_{m_i}\partial_{m_i} F_{\rm TAP}(\bm{m})$ evaluated on the
solution are {\it positive}.
At the mean-field level different local minima are 
separated by infinite barriers therefore the system cannot escape from a local 
minimum in a finite time and hence the minimum (and its basin of attraction) 
is a metastable state of infinite life-time. 
However, despite this simple intuitive picture, not all minima of 
$F_{\rm TAP}(\bm{m})$ can be associated with {\it physical} metastable states 
but only those for which (Plefka's criterion) \cite{Plefka82,Plefka02}  
\begin{equation}
\label{eq:plefka}
 x_{\rm P} = 1 - \frac{c(q)}{N}\,\sum_{i=1}^{N} (1-m_i^2)^2 \geq 0
\end{equation}
where $\beta = 1/T$ and $c(q)$ is a function of $q = (1/N)\sum_i m_i^2$
whose form depends on the interactions. For example for the SK model
$c(q) = \beta^2$ \cite{Plefka82,Plefka02},
while for $p$-spin interaction models 
$c(q) = (\beta^2 p/2)(p-1)q^{p-2}$ \cite{CriLeuRiz03u}.
Therefore within this approach the 
calculation of the number of metastable states is reduced to that of counting
the number of solutions of (\ref{eq:TAPeq}) which are minima and satisfies the
Plefka's criterion (\ref{eq:plefka}) (``physical'' minima).
If it were found that physical minima must satisfy additional constraints,
those must be also included.

Different physical minima may have different free-energy density, 
thus to have a 
better description of metastable states one can group together all 
minima with the same free-energy density and introduce the function
$\rho(f)$ which gives the number of metastable states with 
$F_{\rm TAP}(\bm{m}) = Nf$. 
The configurational entropy is then defined as:
\begin{equation}
\label{eq:ConfEnt}
  \Sigma(f) = \frac{1}{N}\ln \rho(f).
\end{equation}
We are eventually interested into the large $N$ limit, thus 
$\Sigma(f)$ is different from zero only if the number of
physical minima with free energy density $f$ is exponentially large with $N$. 

If we label the 
$\cal{N}_{\rm sol}$ solutions of the TAP equation (\ref{eq:TAPeq}) with 
the subscript $\alpha$ ($\alpha = 1, \ldots, \cal{N}_{\rm sol}$) 
by definition $\rho(f)$ is given by 
\begin{eqnarray}
\label{eq:rhodef}
\rho(f) \stackrel{def}{=}  \sum_{\alpha=1}^{\cal{N}_{\rm sol}}
 \prod_{i=1}^{N} \left[\int\, d m_i\, \theta(\lambda_i^{\alpha})\,
 \delta(m_i - m_i^{\alpha})\right]\,
\nonumber \\
\times \theta(x_{\rm p}^{\alpha})\,\delta\left[F_{\rm TAP}(\bm{m}) - N f\right]
\end{eqnarray}
where $\lambda_i^{\alpha}$ ($i=1,\ldots,N)$ are the eigenvalues of the
Hessian matrix for the $\alpha$-th solution:
\begin{equation}
\label{eq:Hessian}
 \chi_{ij}^{\alpha} = \left. \partial_{m_i}\partial_{m_j}\, 
	F_{\rm TAP}(\bm{m})\right|_{\bm{m} = \bm{m}^{\alpha}}
\end{equation}
and $\theta(x)$ is the Heaviside theta-function. As it stands (\ref{eq:rhodef})
is difficult to handle, however using the properties of delta-function it can 
be transformed into the more manageable form:
\begin{eqnarray}
\label{eq:rho1}
\rho(f) =  \prod_{i=1}^{N} \left[\int\, d m_i\, \theta(\lambda_i)\,
   \delta\left[\partial_{m_i}F_{\rm TAP}(\bm{m})\right]\right]\,
\nonumber \\
\times \det(\underline{\chi}(\bm{m}))\, \theta(x_{\rm P})\,
\delta\left[F_{\rm TAP}(\bm{m}) - N f\right]
\end{eqnarray}
where $\underline{\chi}(\bm{m})$, 
the Hessian matrix (\ref{eq:Hessian}) evaluated 
for a generic $\bm{m}$, is the Jacobian of the transformation  
and $\lambda_i$ are its eigenvalues. The theta-functions ensure that 
the determinant of $\chi$ is always {\it positive} and we have neglected the
the absolute value  of the Jacobian.

In addition to (\ref{eq:rhodef}) we consider the definition without
the theta-functions, which we denote by $\rho_{\rm tot}(f)$, which
counts the total number of TAP solutions.  The effect of the
theta-functions is to eliminate all solutions with at least one negative
eigenvalue, therefore the meaning of $\rho_{\rm tot}(f)$ is not exactly
the same as that of $\rho(f)$ since all solutions are now counted.
There is just one case in which the two formulations, at least in the
limit of our interest, $N\gg 1$, are indeed equivalent: if for large
$N$ the two integral -- with and without theta-functions -- are
dominated by the {\it same set} of solutions.
%  This could not be true
%in general nor for all values of $f$, and hence 
an assumption that must be verified in
each case (and for each value of $f$) separately.

Keeping the sum over all solutions 
is not, however, completely free of difficulties: 
since all solutions are counted the determinant of the Jacobian 
can be negative and the absolute value must be retained
making the subsequent calculation more problematic.
To overcome this difficulties the absolute value is simply dropped
leading to expression:
\begin{eqnarray}
\label{eq:rhocomp}
\tilde{\rho}(f) =  \prod_{i=1}^{N} \left[\int\, d m_i\, 
   \delta\left[\partial_{m_i}F_{\rm TAP}(\bm{m})\right]\right]\,
\nonumber \\
\times \det(\underline{\chi}(\bm{m}))\, 
\delta\left[F_{\rm TAP}(\bm{m}) - N f\right],
\end{eqnarray}
and arguments are given to justify under which circumstances this 
 reproduces the correct result with the absolute value.

We have to compare $\tilde{\rho}(f)$ 
given by (\ref{eq:rhocomp}) with $\rho(f)$.
The question is when $\tilde{\rho}(f)$ yields the same result
as $\rho(f)$. The main difference between (\ref{eq:rho1}) and
(\ref{eq:rhocomp}) is the support of the integrals, larger for
the latter, hence the two expressions are equivalent if the integrals are 
dominated by the {\it same support}. Thus, to extract from $\tilde{\rho}(f)$
the correct result for $\rho(f)$ we should be able to isolate 
the contributions from the common support.
For a generic value of $N$ this could be  quite a hard problem. However,
in the limit of large $N$ where the integrals are evaluated by saddle point 
methods, a simple rule can be applied.

% The contribution 
%of theta-functions is a finite number and hence $\rho(f)$ and 
%$\tilde{\rho}(f)$ are given for $N\gg 1$ by the stationary points of the
%{\it same} functional.

In this case  $\rho(f)$ can be evaluated 
simply considering only the stationary points for which all eigenvalues 
of the Hessian are positive and the Plefka's criterion is satisfied, 
disregarding all others. 
We stress that such constraint is not contained into $\tilde{\rho}(f)$,
so that the functional alone cannot give the desired result.

In the next Sections we shall illustrate this procedure
(re)computing the complexity for the spherical $p$SP-SG model
without external field \cite{CriSom95}
using both expressions (\ref{eq:rhocomp}) and (\ref{eq:rho1}).

\section{TAP Equations for the spherical $p$SP-SG model}
\label{sec:pTAP}
\noindent
The spherical $p$SP-SG model consists of 
$N$ continuous spins $\sigma_i$
interacting via $p$-body interactions \cite{CriSom92}:
\begin{equation}
\label{eq:ham}
  H(\sigma) = \frac{r}{2} \sum_{i=1}^N\, \sigma_i^2 
     - \sum_{1\leq i_1<\cdots<i_p\leq N}\, J_{i_1,\ldots,i_p}\,
       \sigma_{i_1}\cdots\sigma_{i_p} 
\end{equation}
The couplings are quenched independent Gaussian variables with zero mean 
and average $\langle (J_{i_1,\ldots,i_p})^2\rangle = p!/(2 N^{p-1})$. 
The scaling with $N$ ensures a well defined thermodynamic limit. Here and
in the following $\langle(\cdots)\rangle$ denotes disorder average.
The parameter $r$ is a Lagrange multiplier  to impose the global 
constraint $\sum_{i=1}^N\sigma_i^2= N$ on the spins amplitude.

The study of both the static and dynamical properties shows that in
the thermodynamic limit the model presents a (static) transition at a
temperature $T_{\rm s}$, between a high temperature replica symmetric
phase and a low temperature phase with one step of replica symmetry
breaking \cite{CriSom92}.  Despite its simplicity, the spherical
$p$SP-SG model for $p>2$ has an exponentially large number of locally
stable states which dominate the dynamical behaviour above $T_{\rm
s}$.  As a consequence, two-time correlation functions acquire a time
persistent part at a temperature $T_{\rm d} > T_{\rm s}$ which marks
the dynamical transition \cite{CriHorSom93}. The static transition can
be seen as the point where the lowest accessible (metastable) states
dominate.  The dynamical transition, on the contrary, takes place at
the point where the behaviour is ruled by higher, highly degenerate,
metastable states.

The TAP functional has been derived in 
Refs. \cite{KurParVir93,CriSom95}:
\begin{eqnarray}
\beta F_{\rm TAP}(\bm{m}) &=& -\frac{\beta}{p!} 
       \sum_{i_1,\ldots,i_p}\, J_{i_1,\ldots,i_p}\,
       m_{i_1}\cdots m_{i_p} 
\nonumber\\
&&-\frac{N}{2}\ln(1-q)
\nonumber \\
&&-\frac{N\beta^2}{4}\left[1 + (p-1)q^p - pq^{p-1}\right]
\label{eq:pTAPfunc}
\end{eqnarray}
where $Nq = \sum m_i^2$, and taking the derivatives with respect to $m_i$
one obtains the TAP equations. 

The structure of the solutions is better understood performing the change of 
variable $m_i = q^{1/2}\widehat{m}_i$ ($\sum \widehat{m}_i^2 = N$)
which leads to TAP functional density:
\begin{eqnarray}
f_{\rm TAP}(q,E) &=& q^{p/2}\, E -\frac{T}{2}\ln(1-q)
\nonumber \\
&&-\frac{\beta}{4}\left[1 + (p-1)q^p - pq^{p-1}\right]
\label{eq:pTAPqfunc}
\end{eqnarray}
where 
$E = - (1/Np!)\sum J_{i_1,\dots,i_p}\widehat{m}_{i_1}\cdots\widehat{m}_{i_p}$
is the $T=0$ energy density. 
In general, $E$ is a random variable which
depends on both the  realization of couplings and on the orientation of the
vector  $\bm{m}$. However all cases with the same value of
$E$ will also have the same free energy, thus we can 
consider $E$ as given and study the solutions as a function of $E$. 
The TAP equations then reduce to $\partial_{q} f_{\rm TAP}(q,E) = 0$ which 
can be written:
\begin{equation}
\label{eq:sp}
 (1-q)\,q^{p/2-1} = z T
\end{equation}
where 
\begin{eqnarray}
\label{eq:z}
 z &=& \frac{1}{p-1}\left[- E \pm \sqrt{ E^2 - E_{\rm c}^2} \right],
\\
\label{eq:Ec}
 E_{\rm c} &=& -\sqrt{2\,(p-1)/p}.
\end{eqnarray}
It is easy to understand that for any positive $z$ and 
temperature $T$ below
\begin{equation}
\label{eq:Ta}
 T_{\rm a} = (1-q_{\rm a})\,q_{\rm a}^{p/2-1}\, z^{-1}
\end{equation}
where $q_{\rm a} = (p-2)/p$, there are two solutions of the TAP 
equation (\ref{eq:sp}), one larger and one smaller than $q_{\rm a}$.
Anyway, possible candidates for physical solutions are only those which are 
local minima of $f_{\rm TAP}(q,E)$.

By using the TAP equation (\ref{eq:sp}) the second derivative of 
$f_{\rm TAP}(q,E)$ with respect to $q$ evaluated on the solutions 
can be expressed as
\begin{eqnarray}
\partial^2_q\, f_{\rm TAP}(q,E) &=&  \frac{p}{4\beta q}\, 
           \left[q - \frac{p-2}{p}\right]\,
\nonumber\\
	  &&\times \left[\frac{1}{(1-q)^2} - \mu (p-1) q^{p-2}\right]
\nonumber\\
&=&  \frac{p}{4\beta q}\, 
           \left[q - \frac{p-2}{p}\right]\,
	  \frac{z_{\rm c}^2 - z^2}{z_{\rm c}^2}
\label{eq:d2fqTAP}
\end{eqnarray}
where $z_{\rm c} = \sqrt{2/p(p-1)}$ and $\mu = \beta^2 p/2$. 
The requirement of positiveness of the 
second derivative thus selects the solutions
\begin{eqnarray}
\label{eq:qsmall}
q &<& \frac{p-2}{p} \quad \mbox{for}\ z > z_{\rm c}  
\\
\label{eq:qlarge}
q &>& \frac{p-2}{p} \quad \mbox{for}\ z < z_{\rm c}  
\end{eqnarray}
By comparing the two expressions in (\ref{eq:d2fqTAP}) we see that
the condition $z<z_{\rm c}$ is equivalent to
\begin{equation}
\label{eq:eigen}
x_{\rm P} = 1 - \mu (p-1) q^{p-2}\, (1-q)^2> 0
\end{equation}
which also follows from the stability requirement of the replica saddle 
point \cite{CriSom92}
and of the dynamics \cite{CriHorSom93}.
This is the Plefka's criterion (\ref{eq:plefka}) for the physical 
relevance of TAP solutions \cite{Plefka82} for the spherical $p$SP-SG model. 
Indeed it can be easily seen that the  
condition (\ref{eq:qsmall}), for which $x_{\rm P} < 0$,  
leads to an  unphysical $q$ decreasing with temperature.

\section{Complexity of the spherical $p$SP-SG model: standard calculation}
\label{sec:ComppSP}
\noindent
In this Section we report the main steps of the 
calculation of $\tilde{\rho}(f)$ for the spherical $p$SP-SG model. 
Details can be found in the 
literature, see e.g. Refs. \cite{BraMoo80,DedYou83} and
for the specific case of the $p$-spin spin glass model
Refs. \cite{Rieger92,CriSom95,CavGiaPar98}.

The starting point is [cfr. (\ref{eq:rhocomp})]
\begin{eqnarray}
\label{eq:rhopsp}
\tilde{\rho}(f) &=&  N^2\int_{0}^{1}\,dq\, 
                \prod_{i=1}^{N} \left[\int_{-\infty}^{+\infty}\, d m_i\, 
   \delta(G_i)\right]\, \det \underline{A}
\nonumber \\
&\times& \delta\left(Nq - \sum_i m_i^2\right)\,
\delta\left[F_{\rm TAP}(\bm{m}) - N f\right]
\end{eqnarray}
with
\begin{eqnarray}
G_i &=& \partial_{m_i}\beta f_{\rm TAP}(\bm{m}) 
\nonumber\\
    &=&  a(q)\, m_i-
       \frac{\beta}{(p-1)!} \sum_{\bm{j}}\, 
          J_{i,\bm{j}}\, m^{p-1}
\label{eq:G} 
\end{eqnarray}
and
\begin{eqnarray}
A_{ij} &=& \partial_{m_j} G_i
\nonumber\\
       &=& a(q)\, \delta_{ij} - \frac{\beta}{(p-2)!} \sum_{\bm{k}}\, 
          J_{ij,\bm{k}}\, m^{p-2}
\nonumber\\
       &&    + \frac{2}{N}\, a'(q) m_i\,m_j
\label{eq:A}
\end{eqnarray}
where 
\begin{equation}
\label{eq:aq}
  a(q) = \frac{1}{1-q} + \mu (p-1)(1-q) q^{p-2}
\end{equation}
$a'(q) = d a(q)/ dq$ and we have used the short-hand notation:
\begin{equation}
\sum_{\bm{j}}\, J_{i,\bm{j}}\, m^{p-1} \stackrel{def}{=}
\sum_{k_1,\ldots,k_{p-1}}\, 
          J_{i,k_1,\ldots,k_{p-1}}\, m_{k_1}\cdots m_{k_{p-1}}
\end{equation}
and similarly in (\ref{eq:A}). The last term of $A$ is of order $O(1/N)$,
and can be neglected for $N\to\infty$ [see also below].

The structure of the minima is given by the couplings, therefore
$\tilde{\rho}(f)$ (and so $\rho(f)$) may change from sample to
sample. Thus, in principle, to have a well defined complexity we should
introduce replicas to compute $\langle\ln
\tilde{\rho}(f)\rangle$ \cite{BraMoo81}. However it can be
shown \cite{CriSom95,CavGarGia98} that for this model, in absence of a
magnetic field, the annealed average $\ln
\langle\tilde{\rho}(f)\rangle$ is exact, so we can just average
(\ref{eq:rhopsp}) over the disorder.

To perform the average over the couplings it is convenient to use the
integral representation of the delta-function to exponentiate its
argument. This introduces additional parameters which are usually denoted by
$\hat{f}$,  $\hat{q}$ and $\hat{m}_i$ \cite{nota5} 
 conjugated to $f$, $q$
and  $G_i$ and the additional variable $\Delta$ coming from 
Hubbard-Stratonovich transformation.
  The calculation can be further simplified by substituting
$\sum J_{i_1,\ldots,i_p}\, m_{i_1}\cdots m_{i_p}$ from equation
(\ref{eq:G}) in $F_{\rm TAP}(\bm{m})$ [eq. (\ref{eq:pTAPfunc})] and
noticing that the error involved in disorder-averaging the determinant
of $\underline{A}$ separately accounts for changing $\underline{A}$ of
terms of order $O(1/N)$ and hence negligible as
$N\to\infty$ \cite{BraMoo80,Rieger92}.

Performing the averages over the couplings results in
\begin{equation}
\label{eq:rhoave}
\langle\tilde{\rho}(f)\rangle= c \int_{-\infty}^{+\infty} d\hat{f}
                               \int_{0}^{1} dq
                               \int_{-\infty}^{+\infty}\,d\hat{q}\, 
			       \int_{-\infty}^{+\infty} d\Delta\,
                             e^{N\,\Sigma}
\end{equation}
where $c$ is a constant and
\begin{eqnarray}
  \Sigma = i\beta\hat{f}\,\bigl[f-f(q)\bigr]
      + i\hat{q}q - \Delta(1-q) 
\nonumber\\
- \frac{1}{\lambda} \Delta^2 
         + \ln I + G_{x_{\rm p}}(q)
\label{eq:expon}
\end{eqnarray}
with
$f(q)$ the TAP density functional
$f_{\rm TAP}(\bm{m})$ evaluated on the solution of the TAP equation
(\ref{eq:G}):
\begin{eqnarray}
f(q) &=& -\frac{\beta}{4}(1-q^p) 
       -\frac{\beta}{4} (p-2) (1-q) q^{p-1}
\nonumber\\
     &&  -\frac{q\,T}{p\,(1-q)}
       -\frac{T}{2}\ln(1-q),
\end{eqnarray}
\begin{eqnarray}
I &=& \int_{-\infty}^{+\infty}\, \frac{dm\,d\hat{m}}{2\pi}\, 
\exp\Bigl\{
   \frac{\mu q^{p-1}}{2} (i\hat{m})^2 
\nonumber\\
&& 
           + i\hat{m}\left(\frac{1}{1-q}-\Delta\right)m - 
i\hat{q} m^2\Bigr\}.
\label{eq:I}
\end{eqnarray}
and $\lambda = 2\mu(p-1)q^{p-2}$.

The function $G_{x_{\rm P}}(q)$ 
comes from the average of the determinant of $\underline{A}$ which
can be computed either using Grassmann variables \cite{DedYou83} or
introducing replicas \cite{BraMoo80,Rieger92}. 
The form depends on the sign of $x_{\rm P}$ [eq. (\ref{eq:eigen}] 
\cite{Note3}:
\begin{equation}
\label{eq:Gxppos}
G_{x_{\rm P}}(q) = - \ln(1-q), \qquad \mbox{for}\ x_{\rm P} > 0 
\end{equation}
this is $B=0$ solution always adopted in standard calculations  e.g. in
\cite{BraMoo80,Rieger92,CavGiaParMez03}
and
\begin{eqnarray}
G_{x_{\rm P}} &=& \frac{1}{\lambda\,(1-q)^2}\,
               \left[1 - \frac{\lambda^2}{4}\,(1-q)^4\right]
%               \left[1 + \frac{\lambda}{2}\,(1-q)^2\right]
\nonumber\\
	    &&\phantom{xxxx}   + \ln \frac{\lambda}{2}(1-q), \qquad
\mbox{for}\ x_{\rm P} < 0. 
\label{eq:Gxpneg}
\end{eqnarray}
The two expressions coincide for $x_{\rm P}=0$, i.e. for
$\lambda/2 = 1/(1-q)^2$. Details of the calculation can be found
in the Appendix A.

Integration over $\hat{m}, m, \hat{q}, \Delta$ can be
done by the saddle point method \cite{Note1},
which turns out to be exact for the 
integrals over $\hat{m}, m, \Delta$ being Gaussian, while  the 
integral over $\hat{f}$ can be easily performed giving a delta-function.
The final results is then
\begin{eqnarray}
\langle\tilde{\rho}(f)\rangle &\sim& c' \int_{0}^{1} dq\,
                             \delta\bigl[f(q) - f\bigr]\,
                             e^{N\,\Sigma(q)}
\nonumber\\
                         &\sim& e^{N\,\Sigma(q^*)}, 
\qquad N\to\infty
\label{eq:rhoave1}
\end{eqnarray}
where
\begin{eqnarray}
  \Sigma(q) &=& G_{x_{\rm p}}(q)
 + \frac{1}{2} + \frac{1}{2}\ln q - \frac{1}{2}\ln (\mu q^{p-1})
\nonumber\\
&&        +\frac{p-1}{2\mu p q^{p-2}} \left[
              \frac{1}{1-q} - \mu (1-q) q^{p-2}
                                    \right]^2 
\nonumber\\
&& - \frac{1}{2\mu (1-q)^2 q^{p-2}}.
\nonumber\\
\label{eq:sigma}
\end{eqnarray}
and $q^* = q^*(f)$ is the solution of
\begin{equation}
f(q) = f
\label{eq:sol}
\end{equation}
which gives the largest value of $\Sigma(q)$ \cite{Note2}.

The simplest way of studying the solution is  using $q$ as a free parameter 
to scan all values of $f$. This is what is done, for example, in 
Ref. \cite{CriSom95} where the result (\ref{eq:sigma})-(\ref{eq:sol})
was first derived.

\begin{figure}
\includegraphics[scale=0.88]{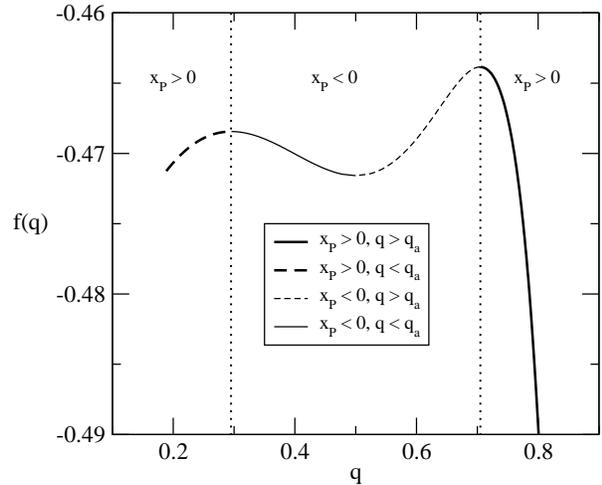}
\caption{Free energy density $f(q)$ as function of $q$ for 
         $p=4$ and temperature $T= 0.51$ between the static transition
         temperature $T_{\rm s} = 0.5030...$ and the dynamical transition
	 temperature $T_{\rm D} = 0.5443...$. 
	 Thicker lines correspond to solutions
         for which the Plefka's criterion is satisfied, while full lines 
	 correspond to solution for which (\protect\ref{eq:d2fqTAP}) is
	  positive, i.e., to local minima of $f_{\rm TAP}(q,E)$.
}
\label{fig:freevq}
\end{figure}

\begin{figure}
\centering
\includegraphics[scale=0.88]{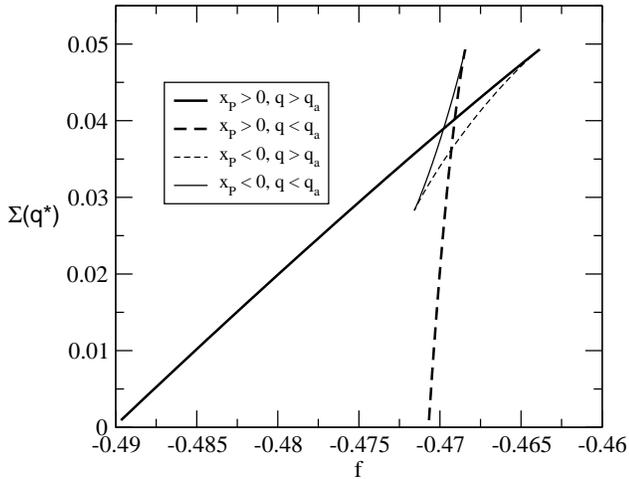}
\caption{$\Sigma(q^*)$ as as function of $f$ for 
         $p=4$ and temperature $T= 0.51$ between the static transition
         temperature $T_{\rm s} = 0.5030...$ and the dynamical transition
	 temperature $T_{\rm D} = 0.5443...$. 
         Thicker lines correspond to solutions
         for which the Plefka's criterion is satisfied, while full lines 
	 correspond to solution for which (\protect\ref{eq:d2fqTAP}) is
	  positive, i.e., to local minima of $f_{\rm TAP}(q,E)$.
         Only values of $f$ for which $\Sigma>0$ are reported.
}
\label{fig:sigmavfree}
\end{figure}

The solutions of eq. (\ref{eq:sol}) can be found using the results of
Section \ref{sec:pTAP}. The free energy 
$f(q)$ as function of $q$ for all stationary points of $f_{\rm TAP}(q,E)$ is
shown in Figure \ref{fig:freevq}  for $p=4$ and temperature $T$
between the static transition temperature $T_{\rm s}$ and the dynamical 
transition temperature $T_{\rm d}$. 
Other values of $p$ or $T$  in this range 
lead to a qualitatively same picture.
The corresponding
$\Sigma(q^*)$ as function of $f$ is shown in Figure \ref{fig:sigmavfree}.

Strictly speaking to evaluate $\langle\tilde{\rho}(f)\rangle$ we should take 
for each value of $f$ the largest value of $\Sigma$ , 
and compute the sign of the neglected coefficient in (\ref{eq:rhoave1}).
However, one is actually interested into the number of metastable states, so
in all calculations done so far all solutions with 
$x_{\rm p}<0$ are cut out, the ``famous'' $B=0$ solution. 
Even if not explicitly stated,
this is in the spirit of the procedure described in Section \ref{sec:Comp}.
We stress, however, that if the procedure is the same the motivations are not.
Indeed the $B\not=0$ solution can also describe minima of the TAP functional
but such configurations violate the Plefka criterion ($x_{\rm p}<0$), 
thus leading to a 
non-physical linear susceptibility.
%is eliminated asserting that it leads to 
%unstable solutions. This is incorrect since the $B\not=0$ solution 
%corresponds to solutions with $x_{\rm p}<0$, among which there are stable
%minima [see, e.g., (\ref{eq:qsmall})]. These (stable) solutions must be
%eliminated since there are unphysical, and not because they are unstable.

\begin{figure}
\centering
\includegraphics[scale=0.88]{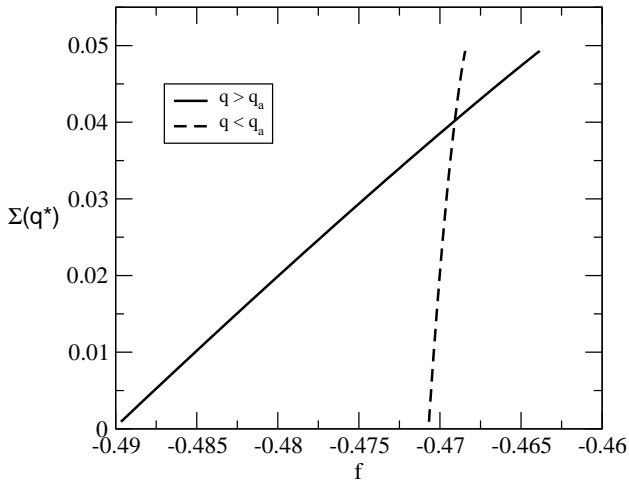}
\caption{Same as Figure \protect\ref{fig:sigmavfree} where only
    solutions satisfying the Plefka's criterion are displayed.
    The full line is the result found in Ref. \protect\cite{CriSom95}.
}
\label{fig:sigmavfreexp}
\end{figure}

If the $x_{\rm p}<0$ solutions are disregarded, we are left with the 
curves shown in Figure \ref{fig:sigmavfreexp} corresponding 
to the solution of the TAP equations with 
$q < (p-2)/p$ (dashed line) and $q > (p-2)/p$ (full line).
Again, if no other information is added, for each $f$
the largest value must be selected to evaluate $\Sigma(f)$.
This means that there is a region of free energies 
where $\Sigma(f)$ is dominated by solutions with $q < (p-2)/p$. 
But these are not local minima of $f_{\rm TAP}$,
see (\ref{eq:d2fqTAP}), and hence for these free energies 
$\Sigma(f)$ does not give the desired result.
This clearly shows that the condition $x_{\rm p}>0$ 
{\it alone} does not guarantees that only 
physical states are counted. To find the correct answer additional
information on the solutions must be added.

For the spherical $p$SP-SG this information is easily obtained. Indeed,
the analysis of the TAP solutions in Section \ref{sec:pTAP}, shows that 
only solutions with $x_{\rm p}>0$ {\it and} $q > (p-2)/p$ do correspond to
metastable states, so only the full line in Figure \ref{fig:sigmavfreexp}
must be considered. This leads to the result first derived in 
Ref. \cite{CriSom95}.

\section{Complexity of the spherical $p$SP-SG model: Hessian Eigenvalues}
\label{sec:HessEig}
\noindent
In the previous Section we have revised step-by-step the standard calculation 
of the complexity for the spherical $p$SP-SG model showing which 
additional information, not included into the definition of
$\tilde{\rho}(f)$, must be added to yield the correct answer.
In this Section we show that that additional information is exactly the 
theta-functions needed to transform $\tilde{\rho}(f)$ into
$\rho(f)$, see eqs. (\ref{eq:rho1}) and (\ref{eq:rhocomp}).

To prove the equivalence we must compute the eigenvalues $\lambda_i$ of the
Hessian matrix (\ref{eq:Hessian}) that for the spherical $p$SP-SG model is
given by (\ref{eq:A}).

The eigenvalues are solutions of the equations
\begin{eqnarray}
\sum_{j}A_{ij}\, \xi_j &=& 
       a(q)\, \xi_i 
      + 2\,q\,a'(q)\, \hat{m}_i\, \frac{1}{N}\, \sum_{j}\hat{m}_j\,\xi_j
\nonumber\\
&& - \frac{\beta\, q^{(p-2)/2}}{(p-2)!} \sum_{j,\bm{k}}\, 
          J_{ij,\bm{k}}\, \hat{m}^{p-2}\,\xi_j 
\nonumber\\
&=& \lambda\,\xi_i
\label{eq:Aeigen}
\end{eqnarray}

There are two classes of eigenvectors $\xi_i$: longitudinal and transversal.

\subsection{Longitudinal Eigenvector}
\noindent
The longitudinal eigenvector is given by:
\begin{equation}
\label{eq:long}
 \xi_i \propto \hat{m}_i, \qquad \forall i
\end{equation}
and hence satisfies the equation:
\begin{eqnarray}
       \left[ a(q) + 2\,q\,a'(q)\right] \, \hat{m}_i\, 
 &-& \frac{\beta\, q^{(p-2)/2}}{(p-2)!} \sum_{\bm{j}}\, 
          J_{i,\bm{j}}\, \hat{m}^{p-1} 
\nonumber\\
&=& \lambda_{\rm L}\,\hat{m}_i
\label{eq:Along}
\end{eqnarray}
Since the Hessian must be evaluated on the solution of the 
TAP equation (\ref{eq:TAPeq}), we can use (\ref{eq:G}) to write 
\begin{equation}
\label{eq:sub}
\beta\, q^{(p-2)/2} \sum_{\bm{j}}\, J_{i,\bm{j}}\, \hat{m}^{p-1} =
  a(q)\, (p-1)!\, \hat{m}_i
\end{equation}
which inserted into (\ref{eq:Along}) leads to
\begin{eqnarray}
\lambda_{\rm L} &=& 2\, q\, a'(q) - (p-2)\,a(q)
\nonumber\\
&=& p
           \left[q - \frac{p-2}{p}\right]
%\nonumber\\
%	  &&\times 
\left[\frac{1}{(1-q)^2} - \mu (p-1) q^{p-2}\right]
\end{eqnarray}
The longitudinal eigenvalue is therefore, apart from positive 
multiplicative coefficients,  
 equal to $\partial^2_q f_{\rm TAP}(q,E)$ evaluated in 
Section \ref{sec:pTAP} [eq. (\ref{eq:d2fqTAP})]. The different coefficients
come from the derivative being taken with respect to $q$ or to $m_i$.
The longitudinal eigenvalue has degeneracy $1$.

Note that the term of  $O(1/N)$ in (\ref{eq:A}) yields a contribution
 of  $O(1)$ for
longitudinal eigenvectors and cannot be neglected \cite{Plefka82}.

\subsection{Transversal Eigenvectors}
\noindent
Transversal eigenvectors satisfy the orthogonality conditions:
\begin{equation}
\label{eq:trans}
 \sum_{i}\xi_i\, \hat{m}_i = 0
\end{equation}
and hence span a space of dimension $N-1$. The eigenvalues equation for
transversal eigenvectors can be written as
\begin{equation}
\label{eq:Atran}
a(q)\, \xi_i - \sum_{j}\widetilde{J}_{ij}\, \xi_j = \lambda_{\rm T}\, \xi_i
\end{equation}

\begin{equation}
\label{eq:jtil}
\widetilde{J}_{ij} = \beta\, q^{(p-2)/p}\, \sum_{k_1<\cdots<k_{p-2}}
              J_{ij,k_1,\ldots,k_{p-2}}\, \hat{m}_{k_1}\cdots
               \hat{m}_{k_{p-2}}
\end{equation}
For large values of $N$ $\widetilde{J}_{ij}$ is a symmetric random
matrix whose elements are independent Gaussian variables with zero average 
and variance:
\begin{equation}
\label{varJt}
\langle(\widetilde{J}_{ij})^2\rangle = \frac{\mu(p-1)q^{p-2}}{N}.
\end{equation}
Therefore for $N\to\infty$ the spectrum of $\widetilde{J}_{ij}$ is
given by the Wigner's semicircular law: \cite{Mehta67,CavGiaPar98},
\begin{eqnarray}
\rho(\lambda_{\rm T}) &=& \frac{1}{2\pi\mu (p-1)q^{p-2}}
\nonumber\\
&&\times
     \sqrt{4\mu (p-1)q^{p-2} - [\lambda_{\rm T} - a(q)]^2}
\label{eq:wigner}
\end{eqnarray}
This gives a spectrum at the leading order in $N$, displaying a non negative
support. Since it can be shown that the tails of this distribution go to zero 
exponentially with $N$ \cite{Mehta67} we can safely exclude negative
eigenvalues.
The thermodynamic limit transversal fluctuations are, thus, 
always stable, regardless of the sign of $x_{\rm P}$ and the whole
stability depends on the longitudinal eigenvalue.
Note, however, that 
the $N-1$ transversal eigenvalues dominate the
calculation of 
\begin{equation}
\det \underline{A} = \exp\left(\mbox{Tr} \ln \underline{A}\right) 
\qquad\mbox{for}\ N\to\infty
\end{equation}
and any information from the longitudinal eigenvalue is washed out
when computing $\tilde{\rho}(f)$.

In conclusion we see that
the procedure described in Section \ref{sec:Comp} 
of selecting the saddle point solutions of $\tilde{\rho}(f)$ 
according to their physical relevance
obtained from the (independent) analysis of the TAP equations produces the 
correct result for the complexity.

\section{Conclusion and Discussion}
\label{sec:Disc}
\noindent
The study of the complex behaviour of glassy systems in terms of the
topological properties of the energy or free-energy surfaces has
recently put new interest into the calculation of the number of
metastable states, also called complexity or configurational entropy,
in mean-field spin glass models.  In this context classical
calculations done for the SK and other disordered spin models have
been reconsidered, extended and also criticised
\cite{CavGiaParMez03,BraMoo03,CriLeuParRiz03,CriLeuParRiz03b}.

Motivated by these controversies in this paper we have reviewed and examined 
further some questions concerning the calculation of the complexity of 
disordered systems. Particular care has been taken to distinguish between
what we would like to compute, $\rho(f)$, and what we are able
to compute, $\tilde{\rho}(f)$. 
We have also discussed how
information on $\rho(f)$ can be extracted, at least in the
thermodynamic limit.
% from $\tilde{\rho}(f)$ by reducing the phase space
%where to look for stationary point of the functional describing
%$\tilde{\rho}(f)$ for $N\to\infty$.

The general approach has been illustrated using the 
spherical $p$SP-SG model, showing the correctness of the reduction procedure.
As by-product we have explicitly shown that the Plefka criterion
separates all solutions of the TAP equations into two classes, both 
containing local minima as well as saddles. 
However, only local minima which satisfy the Plefka criterion 
do represent {\it physical states}.  
The Plefka's criterion is indeed a 
 {\it necessary}, but not sufficient, condition for physical states, and hence
cannot be used alone for the reduction procedure but the requirement of local
stability must be added.

In Ref. \cite{CavGiaParMez03} 
the classical calculation of
Bray and Moore \cite{BraMoo80} and DeDominicis and Young \cite{DedYou83}
for the SK model have been critically revised. 
The main criticism steams from the observation that 
$\hat{\rho}(f)$ (identified with $\rho(f)$ in those papers) can be written as
saddle point calculation over a functional which posses a 
supersymmetry between commuting and anti-commuting variables used to express 
the Jacobian in (\ref{eq:rhocomp}) \cite{ZinnJustin}. 
The classical solution breaks this symmetry and in Ref. 
\cite{CavGiaParMez03} a different, supersymmetric solution was 
proposed. 

In a separate paper \cite{CriLeuParRiz03} we have performed a carefully 
analysis of both the classical and the new supersymmetric solutions. The 
outcome is that both solutions have some limitations.  
For example, the neglected prefactor could be exponentially small in $N$ 
for the classical solution changing its 
prediction \cite{Kurchan91,CriLeuParRiz03}.
On the other hand the supersymmetric solution has a negative 
$x_{\rm p}$, while it is positive for the classical solution.
This point was recently used \cite{BraMoo03} to justify the 
correctness of the classical solution.
However, as we have explicitly shown here for the spherical $p$SP-SG model, 
this condition is not a sufficient condition. 
In order to prove the correctness of the solution one should prove that
it corresponds to a physical (stable) state. A rather difficult 
problem already at the annealed (replica symmetric) level used in these 
calculations. Moreover, since it is known that a (marginally) stable solution
for the SK model requires an infinite-number of replica symmetry breakings,
the requirement of local stability may partially
or totally wash-out the results from the annealed approximation.
A complete calculation of the complexity of the SK model must include
full replica-symmetry breaking, making not only the calculation but
also the analysis of the saddle points more
difficult \cite{BraMoo81,CriLeuParRiz03b,AnnCavGiaPar03}.
Supersymmetry requirements
introduce partial simplifications, since they lead to a connection
between the complexity and the replica calculation.  However, we
stress that supersymmetry it is not an {\it a-priori} requirement for
the complexity.  Indeed, while $\tilde{\rho}(f)$ is supersymmetric, the
reduction procedure needed to go from $\tilde{\rho}(f)$ to $\rho(f)$ may
destroy the supersymmetry, so that supersymmetry must be proved case
by case.  It turns out that in the case of the $p$SP-SG model
discussed here the supersymmetry is conserved \cite{CavGarGia98}, a
property which can be associated with the fact that all metastable
states are uniquely identified by its energy at $T=0$ and hence its
number is conserved \cite{KurParVir93}.

Indeed using the TAP equation
(\ref{eq:sp}) $\Sigma(q)$ for the spherical $p$PS-SG can be rewritten
as:
\begin{eqnarray}
\Sigma^{+}(z) = \frac{1}{2}\left[
            \frac{2-p}{p} - \ln\frac{pz^2}{2} + \frac{p-1}{p} z^2 
            - \frac{2}{p^2z^2}
                       \right]
\label{eq:sigposz}
\end{eqnarray}
for $x_{\rm P} > 0$ \cite{CriSom95}, 
and 
\begin{eqnarray}
\Sigma^{-}(z) &=& \Sigma^{+}(z) 
            + \ln\left[\frac{p(p-1)}{2}z^2\right]
\nonumber\\
&&  + \frac{1}{p(p-1)z^2} \left[1 - \frac{p^2(p-1)^2z^4}{4}\right]
\label{eq:signegz}
\end{eqnarray}
for $x_{\rm P}<0$.
Such  property is in agreement with the argument presented in Sec. I.D of Ref.
\cite{CriLeuParRiz03}
 where the equivalence between supersymmetry 
and  stability of the TAP equations under external field perturbation
is shown.

\appendix
\section{Appendix}
\noindent
Here we calculate $\langle \det\underline{A}\rangle$ for $N\to\infty$ 
using the identity [see, e.g., Ref. \cite{ZinnJustin}]
\begin{equation}
\label{eq:grass}
 \det\underline{A} = \int\prod_{i=1}^{N}\, d\eta_i\,d\eta^+_i\,
   \exp\left(\sum_{ij} \eta^+_i A_{ij}\eta_j\right)
\end{equation}
where $\eta_i$ and $\eta^+_i$ are anti-commuting (Grassmann) variables.
From eq. (\ref{eq:A}) we have
%\begin{widetext}
\begin{eqnarray}
\langle\det\underline{A}\rangle &=&
 \int\prod_{i=1}^{N}\, d\eta_i\,d\eta^+_i\, 
          \exp\left(a(q) \sum_{i} \eta^+_i \eta_i\right)
\nonumber\\
&&\times   \prod_{i_1<\cdots<i_p} \left\langle
     \exp\left[-\frac{\beta}{(p-2)!} J_{i_1,\ldots,i_p} \right.\right.
\nonumber\\
&& \phantom{xxxx}\times\left.\left.           \sum_{\pi} 
      \eta^+_{\pi(i_1)} \eta_{\pi(i_2)} m_{\pi(i_3)}\cdots m_{\pi(i_p)}
         \right]
       \right\rangle
\nonumber\\
&=&
 \int\prod_{i=1}^{N}\, d\eta_i\,d\eta^+_i\, 
          \exp\left(a(q) \sum_{i} \eta^+_i \eta_i\right)
\nonumber \\
&& \times
   \prod_{i_1<\cdots<i_p} 
     \exp\left[\frac{\mu}{2N^{p-1}} \frac{p!(p-1)!}{(p-2)!^2} 
\right.
\nonumber\\
&&\phantom{xxx}\times           \sum_{\pi} 
      \eta^+_{\pi(i_1)} \eta_{\pi(i_2)} m_{\pi(i_3)}\cdots m_{\pi(i_p)}
\nonumber\\
&&\phantom{xxxxxxxxxxxxxxx}\times
      \eta^+_{i_1} \eta_{i_2} m_{i_3}\cdots m_{i_p}
         \Bigr]
\label{eq:detave}
\end{eqnarray}
%\end{widetext}
where $\sum_{\pi}$ is a sum over all permutations of $p$ different integers
$i_1,\ldots,i_p$. When the products are expanded only terms which contain
pairs of $\eta^+_i\eta_i$ with the same index survive. Since since there are
$(p-2)!$ terms with the same pairs of Grassmann variables, we end up with
%\begin{widetext}
\begin{eqnarray}
\langle\det\underline{A}\rangle &=&
 \int\prod_{i=1}^{N}\, d\eta_i\,d\eta^+_i\, 
          \exp\left(a(q) \sum_{i} \eta^+_i \eta_i\right)
\nonumber\\
&&\times 
     \exp\left[\frac{\mu}{2N^{p-1}} \frac{p!(p-1)!}{(p-2)!} 
   \right.
\nonumber\\
&&\phantom{xxxxxx}\times  \left.
          \sum_{i_1<\cdots<i_p} 
      \eta^+_{i_1} \eta_{i_1} \eta^+_{i_2} \eta_{i_2} 
      m^2_{i_3}\cdots m^2_{i_p}
         \right]
\nonumber\\
&=& 
 \int\prod_{i=1}^{N}\, d\eta_i\,d\eta^+_i\, 
          \exp\left[a(q) \sum_{i} \eta^+_i \eta_i \right.
\nonumber\\
&& \phantom{xxxxxx}
\left.
 +  \frac{\mu(p-1)}{2N} q^{p-2}\left(\sum_i\eta^+_{i} \eta_{i}\right)^2
         \right]
\label{eq:detave1}
\end{eqnarray}
%\end{widetext}
where we have used $Nq = \sum_{i} m_i^2$. The square in the exponential can be 
open using a Stratonovich-Hubbard transformation. The resulting expression is
diagonal in $\eta^+_i\eta_i$ and the integral over the Grassmann variable 
can be easily done. After a simple algebra we get
\begin{eqnarray}
\langle\det\underline{A}\rangle &=& 
  \int_{-\infty}^{+\infty}\frac{dz}{\sqrt{2\pi\sigma^2(q)/N}}
\nonumber\\
&&\times \exp N\left[ -\frac{z^2}{2\sigma^2(q)} + \ln(a(q)+iz)\right]
\label{eq:detave2}
\end{eqnarray}
where $\sigma^2(q) = \mu(p-1)q^{p-2}$.
Finally performing the change of variable
$iz + (1-q)\sigma^2(q) = i x$ we end up with \cite{Note4}:
\begin{equation}
\langle\det\underline{A}\rangle =
\frac{\exp{\left(\frac{\sigma^2(q)(1-q)^2}{2}\right)}}
{\sqrt{2\pi\sigma^2(q)/N}} 
  \int_{-\infty}^{+\infty} dx
  e^{NG(x)}
\end{equation}
\begin{equation}
\label{eq:sap}
G(x) = -\frac{x^2}{2\sigma^2(q)} - ix(1-q) + \ln\left(\frac{1}{1-q}+ix\right)
\end{equation}
For $N\to\infty$ the integral can be done by saddle point method:
\begin{equation}
\frac{dG(x)}{dx} = x\left[\frac{(1-q)^2}{1+ix(1-q)} - \frac{1}{\sigma^2(q)}
                    \right] = 0
\end{equation}
which admits two solutions:
$x=0$ and $x\not=0$. 
Stability requires that the saddle point be a maximum:
\begin{equation}
\frac{d^2G(x)}{dx^2} = \frac{(1-q)^2}{[1+ix(1-q)]^2} - \frac{1}{\sigma^2(q)}
                     < 0
\end{equation}
For the $x=0$ solution this implies that [cfr. eq. (\ref{eq:eigen})]
\begin{equation}
x_{\rm P} = 1 - \sigma^2(q)(1-q)^2 > 0
\end{equation}
and $G(x)$ reduces to (\ref{eq:Gxppos}).

It is easy to see that the $x\not=0$ solution is stable only if
\begin{equation}
x_{\rm P} = 1 - \sigma^2(q)(1-q)^2 < 0
\end{equation}
in which case $G(x)$ reduces to (\ref{eq:Gxpneg}).


\begin{thebibliography}{99}
\bibitem{nota5}
  Not to be confused with the angular part of $m_i$ used in Sec. III. We 
leave it like this because it is the standard notation in literature.
\end{thebibliography}
\end{document}